\documentstyle[12pt]{article}

\begin{document}

\begin{titlepage}
\begin{flushright}
{\large \bf UCL-IPT-96-22}
\end{flushright}
\vskip 1.9cm
\begin{center}

{\Large \bf
Effective $SU(2)_L \otimes U(1)$ theory and the Higgs boson mass}
\vskip 1cm

{\large Zhen Yun Fang$^a$, 
G. L\'{o}pez Castro$^{b,\ }$\footnote{Permanent
 address: Departamento de F\'{i}sica, CINVESTAV del IPN, Apartado Postal
14-740, 
07000 M\'{e}xico D.F., M\'{e}xico}, J. L. Lucio M.$^c$  and  J. 
Pestieau$^b$}\\

 {\em $^a$ Department of Applied Physics, Chongqing University, 
Chongqing,}\\ {\em Sichuan 630044, People's Republic of China} \\

{\em $^b$ Institut de Physique Th\'eorique, Universit\'e catholique de
Louvain,}\\
{\em  B-1348 Louvain-la-Neuve, Belgium}

{\em $^c$ Instituto de Fisica, Universidad de Guanajuato, Apartado}\\
{\em Postal E-143, 37150 Le\'on, Guanajuato, M\'exico}
 
\end{center}

\vskip 1.6cm

\begin{abstract}
  We assume the stability of vacuum under radiative corrections in the 
context of the standard electroweak theory. We find that this theory 
behaves as a good effective model already at cut off energy scales as low 
as 0.7 TeV. This stability criterion allows to predict $m_H= 318 \pm 13$ 
GeV for the Higgs boson mass.
\end{abstract}

\


\end{titlepage}%

\medskip

  Since the standard electroweak theory is renormalizable, the 
ultraviolet (quadratic and logarithmic) divergencies appearing in loop 
calculations can be removed by proper redefinitions of a small number of 
parameters ({\em e.g.}, masses). The renormalization program allows to 
get finite predictions for physical quantities except for those few 
parameters which require redefinitions. In this way, we renounce to 
give some physical content to the divergent behaviour of the 
theory at very high energies \cite{chanowitz}.

  Instead of following the renormalization program, we can consider that 
$SU(2)_L \otimes U(1)$ is an {\em effective} theory with an ultraviolet 
cut off at the energy scale $\Lambda$ and try to trigger some information 
on the model itself from the behaviour of divergent terms. In other 
words, we assume that a more complete theory than $SU(2)_L \otimes U(1)$ 
introduces new physical effects at high energy that cut off the 
ultraviolet divergencies. Below $\Lambda$, $SU(2)_L \otimes U(1)$ is 
supposed to describe all the electroweak processes in a satisfactory way.

  In this letter we consider $SU(2)_L \otimes U(1)$ as an effective 
theory below $\Lambda$ and assume the vanishing of the common divergent 
contributions (tadpoles) to the masses of all the particles.
We find that the electroweak 
theory becomes effective even at cut off energies as small as $\Lambda 
\sim 0.7$ TeV. This criterion for vacuum stability under radiative 
corrections allows to derive a mass for the Higgs boson that differs from 
its asymptotic value ($\Lambda 
\rightarrow \infty$) $m_H=\sqrt{4(m_t^2+m_b^2)-2m_W^2-m_Z^2}$ \cite{roger}-
\cite{veltman} by less than 0.6 \% already at $\Lambda \sim 0.7$ TeV.

  Let us start by considering the divergent one-loop tadpole contributions. 
These 
tadpoles give a universal contribution to self-mass corrections because 
the bare masses of all the particles are proportional to the vacuum 
expectation 
value $v$ of the Higgs field. The tadpoles are gauge-dependent and are 
given by \cite{fleischer}
\begin{eqnarray}
\frac{\delta v_t}{v} &=& \left( \frac{\alpha}{16 \pi} \right) 
\frac{1}{m_H^2 m_W^2 \sin^2 \theta_W} \left \{ -2m_Z^4-4m_W^4 
-\frac{3}{2} m_H^2 A_0(m_H) -3m_Z^2A_0(m_Z) \right. \nonumber 
\\ && \hspace{2.5cm}\left. -6m_W^2A_0(m_W) 
-\frac{1}{2}m_H^2A_0(\sqrt{\xi}m_Z)-m_H^2A_0(\sqrt{\xi}m_W)
\right. \nonumber \\
&& \hspace{2.5cm} \left. +12[m_t^2A_0(m_t)+m_b^2A_0(m_b)] \right\}
\end{eqnarray}
where $m_i\ (i=W,Z,H,t,b)$ denote the masses of gauge bosons, the Higgs 
boson and top and bottom quarks (we have neglected fermion masses other 
than $m_{t,b}$), $\xi$ is the gauge parameter ($\xi=1$ in the 
t'Hooft-Feynman gauge and $\xi=0$ in the Landau gauge) and \cite{fleischer}

\begin{equation}
A_0(m)= \left\{
\begin{array}{lc}
-m^2\left ( \frac{\textstyle 2}{\textstyle \epsilon} -\gamma + \log 
(4\pi\mu^2) + 1 -\log m^2 \right ), \\
\Lambda^2 -m^2\log \frac{\textstyle \Lambda^2}{\textstyle m^2} -m^2 
\end{array} \right.
 \end{equation}
The two expressions in Eq. (2) are obtained by using the $n$-dimensional 
and corresponding  cut off regularization methods, respectively.

In a similar way, we can compute the corrections to the neutrino mass
 (we assume a non-vanishing value for the neutrino mass). We get
\begin{eqnarray}
\frac{\delta m_{\nu}}{m_{\nu}} &=&\left( \frac{\alpha}{16 \pi} \right)
\frac{1}{ m_W^2 \sin^2 \theta_W} \left\{ \frac{3}{2} 
m_W^2+B_0(0,0,\sqrt{\xi}m_W) \right. \nonumber \\
&&\hspace{2.5cm}Ê\left. +\frac{3}{4} 
m_Z^2+\frac{1}{2}B_0(0,0,\sqrt{\xi}m_Z) \right\} + \frac{\delta v_t}{v}
\end{eqnarray}
where \cite{fleischer} 
\begin{equation}
B_0(0,0,m) =
\left\{
\begin{array}{lc}
-m^2\left ( \frac{\textstyle 2}{\textstyle \epsilon} -\gamma + \log 
(4\pi\mu^2) + 1 -\log m^2 \right ), \\
 -m^2\log \frac{\textstyle \Lambda^2}{\textstyle m^2} -m^2,
\end{array} \right.
\end{equation}
respectively, using the $n$-dimensional and cut off methods. Eq. (3) is 
in agreement with corresponding results in Ref. \cite{drell} where finite 
terms of the form $m^2$ and $m^4$ are neglected.

  As it can be easily checked from the previous expressions, $\delta 
m_{\nu}/m_{\nu}$ is explicitly gauge-independent. This means that if we 
write the tadpoles in the Landau gauge we obtain 
\begin{equation}
\left. \frac{\delta v_t}{v} \right |_{\xi=0} = \frac{\delta 
m_{\nu}}{m_{\nu}} - \frac{3}{4} \left( \frac{\alpha}{16 \pi} \right)
\frac{1}{ m_W^2 \sin^2 \theta_W}
\{2m_W^2+m_Z^2 \},
\end{equation}
{\em i.e.}, this quantity is $SU(2)_L \otimes U(1)$ gauge-invariant. As is 
well 
known, the tadpoles can be computed from an effective potential only 
in the Landau gauge \cite{weinberg}.

   Now let us assume that tadpole contributions in the Landau gauge 
vanish, namely $\delta v_t =0$. According to Eq. (5), $\delta 
m_{\nu}/m_{\nu}$ becomes negligible small ($\sim 1.6 \times 10^{-3}$), 
{\em i.e.} the radiative corrections to the small neutrino mass are very 
small. Since tadpole radiative corrections affect the masses of all the 
particles in an universal way, the stability of the vacuum under 
radiative corrections ($\delta v_t =0$) implies that the masses of all 
the particles in the effective model scale as $v_{class}$, the classical 
vacuum expectation value of the Higgs boson.

  In figure 1, we plot the Higgs boson mass as a function of $\Lambda$ 
when we imposse the condition $\delta v_t=0$ in the Landau gauge. The three 
different curves correspond to $m_t= 169,\ 175$ and $181$ GeV and we have 
used $m_W=80.35$ GeV, $m_Z=91.187$ GeV and $m_b = 4.5$ GeV \cite{pdg}.
  When $\Lambda > 0.7$ TeV, we observe that the mass of the 
Higgs boson is approximately given by the relation 
\begin{eqnarray}
m_H^2 &=& 4(m_t^2+m_b^2)-2m_W^2-m_Z^2  \\
 &\approx& 318 \pm 13\ {\rm GeV}.
\end{eqnarray}
Eq. (6) corresponds to the condition for the cancellation of quadratic 
divergencies \cite{roger}-\cite{veltman} in the self-masses of {\em all} the 
particles in the standard electroweak theory.

 If we 
have chosen $\delta m_{\nu}/m_{\nu}=0$ instead of $\delta v_t/v=0$, the 
plots obtained would have been almost 
identical for $\Lambda \geq 0.7$ TeV because finite terms in Eq. (5) are 
negligible above these cut off energies. It is interesting to note that 
our predicted value for $m_H$ is somewhat higher than the one predicted 
in another version of the effective electroweak theory \cite{hambye} 
where the Higgs potential is reduced to its quartic term.

  In conclusion, by assuming the vanishing of tadpole radiative 
corrections we find that the standard electroweak theory behaves as a 
good effective model with a relatively small cut off energy scale. The 
mass obtained for the Higgs boson departs by less than 0.6 \% from the 
value obtained by requiring cancellation of quadratic divergencies 
\cite{roger}-\cite{veltman} already at $\Lambda \sim 0.7$ TeV. Thus, no 
fine-tunning of physical masses \cite{chanowitz} are required to suppress 
the quadratic divergent 
corrections to self-masses in the effective version of the standard 
electroweak theory.

\bigskip

\noindent {\bf Acknowledgements}

\medskip

   We thank Abdelhouahed Bernicha, Jean-Marc G\'erard, Thomas Hambye, 
Mat\'\i as Moreno and Jacques Weyers for valuable discusions.

\newpage

\medskip

\begin{center}
FIGURE CAPTION
\end{center}

Figure 1: Mass of the Higgs boson as a function of $\Lambda$ by imposing 
$\delta v_t=0$. The plotted curves (from bottom to top)  correspond to 
$m_t=169,\ 175$ and $181$ GeV, respectively. Units in both axis are TeV.

\newpage



\begin{thebibliography}{9}
\bibitem{chanowitz} See for example: M. Chanowitz, {\em Ann. Rev. Nucl. 
Part. Sci.} {\bf 38} (1988) 323.

\bibitem{roger} R. Decker and J. Pestieau, preprint UCL-IPT-79-19; DESY 
Workshop, October 22-24, 1979; {\em Mod. Phys. Lett.} {\bf A7} (1992) 
3773; G. L\'opez Castro and J. Pestieau, {\em Mod. Phys. Lett} {\bf A10} 
(1995) 1155.

\bibitem{veltman}  M. Veltman, {\em Acta Phys. Polonica} {\bf B12} 
(1981) 437.

\bibitem{fleischer} J. Fleischer and F. Jegerlehner, {\em Phys. Rev.} 
{\bf D23} (1981) 2001.  

\bibitem{drell} I.-Hsiu Lee and  S. D. Drell, in: {\sl M.A.B. B\'eg Memorial 
Volume}, Eds. A. Ali and P. Hoodbhoy, (World Scientific,Singapore, 1991) 
p.13. 

\bibitem{weinberg} S. Weinberg, {\em Phys. Rev.} {\bf D7} (1973) 2887.

\bibitem{pdg} {\sl Review of Particle Properties}, {\em Phys. Rev.} {\bf 
D54} (1996) Part I.

\bibitem{hambye} T. Hambye, {\em Phys. Lett.} {\bf B371} (1996) 87.
\end{thebibliography}
 
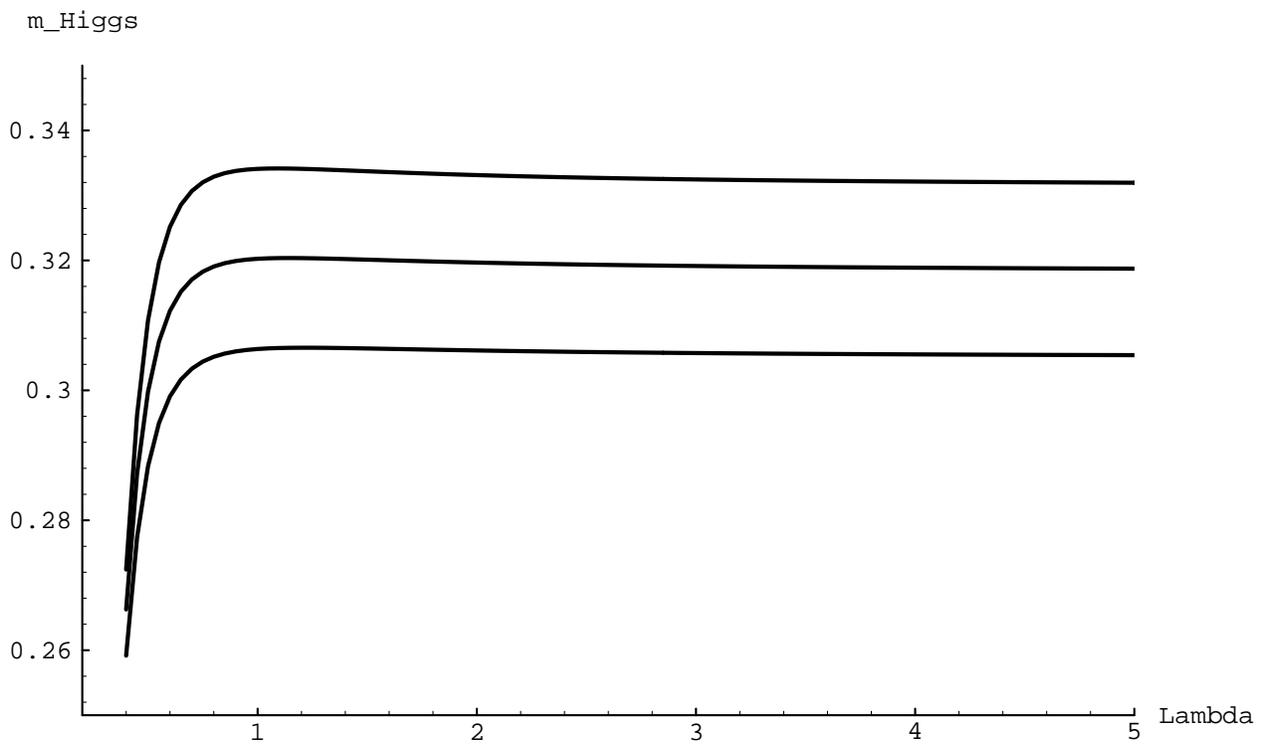
\end{document}